\documentclass[11pt]{article}

\usepackage[a4paper,margin=1in]{geometry}
\usepackage{amsmath,amssymb,amsfonts}
\usepackage{graphicx}
\usepackage{bm}
\usepackage{graphicx}
\usepackage{cite}
\usepackage{url}
\usepackage{enumitem}
\usepackage{siunitx}
\usepackage{hyperref}
\usepackage{authblk}
\title{X-ray dark-field imaging from intensity flow: A Fokker--Planck approach to grating interferometry}

\author[1]{Samantha J.~Alloo}
\author[2]{Florian~Schaff}
\author[3]{Regine~Gradl}
\author[2]{Benedikt~G{\"u}nther}
\author[2,4,5]{Franz~Pfeiffer}
\author[1]{Kaye~S.~Morgan}

\affil[1]{School of Physics and Astronomy, Monash University, Australia}
\affil[2]{Department of Physics, TUM School of Natural Sciences, Technical University of Munich, Germany; Munich Institute of Biomedical Engineering, Technical University of Munich, Germany}
\affil[3]{Institute of Biomedical Physics, Medical University of Innsbruck, Austria}
\affil[4]{Institute for Diagnostic and Interventional Radiology, TUM School of Medicine and Health, TUM Klinikum, Germany}
\affil[5]{TUM Institute for Advanced Study, Technical University of Munich, Germany}

\date{} 

\begin{document}

\maketitle

\begin{abstract}
Grating interferometry is a promising diagnostic technique that enables simultaneous acquisition of three complementary, synergistic X-ray images: transmission, differential phase, and dark-field. A key advantage of grating interferometry over other setups is its ability to use large pixels and, hence, large-area detectors, as well as its compatibility with low-coherence, compact X-ray sources, both of which are key factors for human-scale imaging. It has already demonstrated strong potential for chest imaging applications, including the diagnosis of pulmonary emphysema, fibrosis, and cancer. To retrieve transmission, differential phase, and dark-field images from the intensity data, a dedicated algorithm is required to separate the distinct physical mechanisms contributing to the measured contrast. Since its first realization, this image-retrieval step has remained fundamentally unchanged. In this work, we develop a novel transmission- and dark-field retrieval algorithm for grating-interferometry data derived from the X-ray Fokker--Planck equation. To demonstrate and validate our Fokker--Planck algorithm, we apply it to experimental measurements of a test sample with known materials and to data from a mouse chest acquired with varying exposure times and with added Poisson noise. The retrieved images were qualitatively and quantitatively compared with those retrieved using a conventional sinusoidal-fitting approach. Across both samples, the Fokker--Planck method produced results consistent with conventional image retrieval, with a comparable signal-to-noise ratio. Notably, our Fokker–Planck method suppresses artefacts arising in the conventional approach under grating perturbations (e.g., structural defects such as scratches) and reduced flux or visibility, yielding smoother and more reproducible images. Additionally, we demonstrate that our Fokker--Planck method has an advantage over the conventional dark-field retrieval method for fast sample imaging with short exposure times and high noise. This improvement arises because the Fokker--Planck framework incorporates information from neighboring pixels, while the conventional single-pixel approach is more susceptible to perturbations and pixel-wise intensity variations. The presented algorithm can be easily integrated into existing pipelines, with an open repository link provided.
\end{abstract}


\section{Introduction}
\begin{figure}[ht] 
    \centering
    \includegraphics[width=0.8\textwidth]{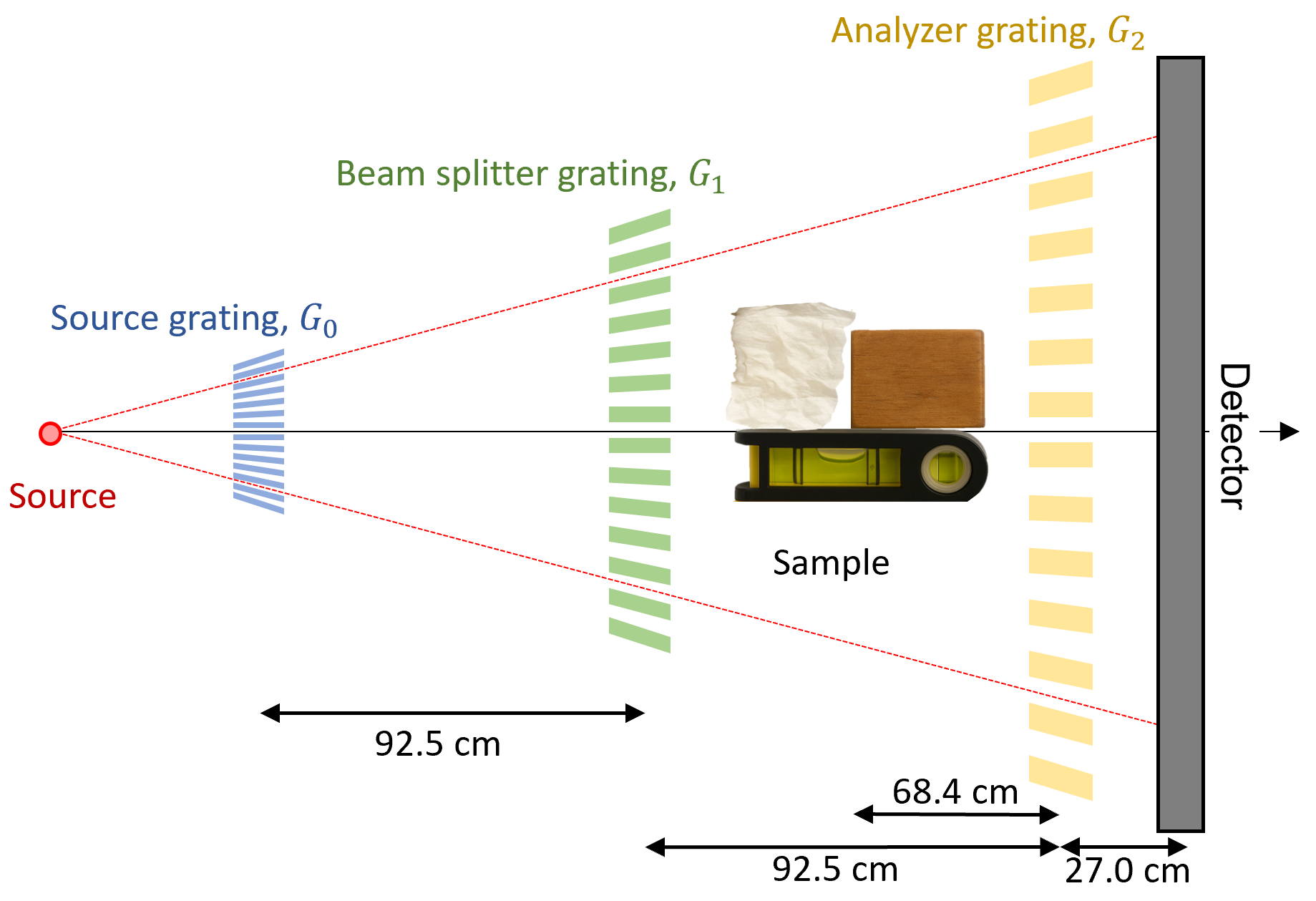}
    \caption{\textit{Grating interferometry (GI) imaging setup established at the Technical University of Munich (TUM) for imaging of the test sample. This sample comprised a piece of wood (brown) and scrunched tissue paper (white) placed on a spirit level (black with yellow interior).}}
    \label{fig:Setup}
\end{figure}

X-ray phase and dark-field effects can be exploited to extract additional structural information beyond what is accessible in conventional absorption imaging, which only considers the transmitted X-ray intensity. Phase contrast helps distinguish low-density materials that weakly attenuate X-rays and are therefore almost invisible in conventional absorption images. Dark-field contrast is generated by sample regions with unresolved microstructure or multiple interfaces, which scatter X-rays and cause local diffusion of X-ray intensity downstream of the sample. Structures that are invisible in absorption and phase imaging, because they are too small to be resolved, are mapped in dark-field images. 
\\\\
Several experimental methods exist for measuring X-ray phase and dark-field. This work focuses on grating interferometry (GI) \cite{Momose2003,Weitkamp2005,Pfeiffer2008}, commonly implemented as the Talbot-Lau interferometer, a technique demonstrated at synchrotron X-ray sources as well as at compact sources, including conventional X-ray tubes \cite{pfeiffer2006} and inverse-Compton sources \cite{Bech2009,Jud2017}. Because GI does not need to resolve phase and dark-field effects directly, it does not require a detector with small pixels and is therefore suitable for imaging large objects. Due to this attractive feature, GI has been widely applied in research and industry-related studies. GI has become a particularly promising tool for diagnostic imaging \cite{Momose2014}, with dark-field chest X-ray imaging demonstrated in clinical studies \cite{Willer2021,Urban2023} and the first clinical dark-field computed tomography prototype in development \cite{Viermetz2022}. For a comprehensive and recent review of GI and developments, see Wang, 2024 \cite{wang2024recent}.
\\\\
GI at a conventional X-ray source uses three gratings: a source grating $G_0$ to create an array of mutually coherent sources of X-rays (noting this is not required for coherent X-ray sources like a synchrotron), a beam splitter grating $G_1$ to create an interference pattern, and an analyzer grating $G_2$ to study the interference pattern; Fig.~\ref{fig:Setup} shows one of the GI setups used in this work. Via the Talbot effect, $G_1$ generates a 1D periodic fringe pattern with intensity modulation perpendicular to the grating lines, and in the presence of a sample, this pattern will be locally modified according to the sample's structural characteristics. As the fringe period is typically smaller than the size of the detector's pixels, an absorbing analyzer grating $G_2$, matched in pitch to the fringe's pitch, is introduced to convert the fringe patterns into measurable intensity variations. Though single-shot methods exist, such as Bevins \textit{et al.} \cite{bevins2012multicontrast} (with limitations in spatial resolution and visibility), the typical approach to separate transmission, phase, and dark-field effects is to acquire a `phase-stepping curve' in both the absence and presence of the sample. During a scan, the analyzer grating $G_2$ is moved laterally in equidistant steps across a full fringe pitch, with an image captured at each step. The sequence of images acquired for different grating positions in the absence of the sample is known as the reference scan, while data acquired with the sample in the beam is the sample scan. The core idea of this acquisition procedure is that each pixel's intensity variation, measured across grating steps over a full period, can be modeled as a sinusoid. Relative to an empty beam, for a given pixel, the sinusoid is reduced in mean value when there is X-ray absorption in the sample, transversely shifted by phase variations, and its relative amplitude or visibility is reduced when there is scattering from unresolved microstructures or multiple interfaces, measured as a dark-field signal. These changes can be quantified using either sinusoidal fitting \cite{Weitkamp2005,Pfeiffer2008} or Fourier analysis \cite{weber2011noise} to extract the parameters of the sinusoid for the reference and sample scans. By extracting these sample-induced changes to the stepping curve pixel-by-pixel, the sample's transmission, phase, and dark-field images can be retrieved.
\\\\
Although imaging via GI was realized almost two decades ago, and there have been, and continue to be, significant developments in instrumentation \cite{arboleda2017sensitivity,pereira2025quantifying} and applications \cite{Ludwig2018,hellerhoff2025grating}, the process of image retrieval has remained largely unchanged. Pixel-wise fitting of phase-stepping curves has been the gold-standard image-retrieval method in GI. Improvements in retrieved image quality typically arise from pre-processing the data, such as correcting the phase-stepping positions \cite{Kaeppler2017}; from mathematical refinements during retrieval \cite{Makinen2025}, including the use of more advanced fitting models beyond the conventional sinusoidal \cite{noichl2023correction}; or from optimization of experimental conditions \cite{spindler2025simulation}. A recent development is patch-wise image retrieval, which incorporates information from neighboring pixels \cite{wirtensohn2024self}. 
\\\\
In this work, we propose a fundamentally new approach to retrieving transmission and dark-field images from GI data, exploiting the X-ray Fokker--Planck equation (XFPE) \cite{paganin2019x, morgan2019applying}. The XFPE models image contrast formation in terms of coherent and diffusive energy flows, determined by the object's phase and dark-field properties, respectively. This equation can be viewed as a diffusive (dark-field) extension of the widely adopted transport-of-intensity equation (TIE) \cite{teague1983deterministic}, which has successfully provided a model for phase recovery across various imaging modalities and wavelengths \cite{mitome2021transport, zuo2020transport}, most notably in the well-known Paganin method for propagation-based X-ray phase-contrast imaging \cite{paganin2002simultaneous}.  X-ray phase and dark-field retrieval via the XFPE was realized just five years ago and has since been successfully applied within propagation-based X-ray imaging \cite{leatham2023x, leatham2024x, ahlers2024x, ahlers2025single} and speckle/single-grid-based X-ray imaging \cite{pavlov2020x, pavlov2021directional, alloo2022dark,alloo2023m, beltran2023fast,alloo2025separating, liu2025low}. These experimental approaches are best suited to high-resolution X-ray imaging, as they directly resolve phase and dark-field effects. For example, in speckle/single-grid-based imaging, the X-ray illumination is patterned using sandpaper, a grid, or a grating, and this pattern is directly resolved in the detector-acquired reference image. When the sample is introduced, the distortions in the reference pattern are also resolvable, and these distortions reveal the sample's properties: transverse shifts in the pattern relate to phase effects, and local blurring indicates dark-field effects. In propagation-based and speckle-based X-ray imaging, XFPE-based phase retrieval algorithms have demonstrated the ability to retrieve a sharper and more accurate phase image than TIE-based algorithms, since image blur from dark-field effects is separated out through the simultaneous retrieval of the dark-field image \cite{leatham2024x, alloo2024stabilizing}. Notably, XFPE-derived algorithms retrieve dark-field images that are smooth and robust to noise \cite{leatham2023x,ahlers2024x,beltran2023fast}. This property may be seen because the images are recovered via linear operations on entire images, meaning information is incorporated not only from neighboring pixels but across the entire image domain. This differs from dark-field retrieval using alternative `local' methods, such as convolution-based retrieval \cite{how2022quantifying} or Unified Pattern Modulation Analysis (UMPA) \cite{zdora2017x} in speckle/grid-based imaging, visibility approaches in propagation-based imaging \cite{ahlers2024x}, or conventional sinusoidal fitting approaches used in GI \cite{strobl2014general, Wang2020}. These approaches use local analysis windows ranging from one to several pixels in area to quantify how the recorded data (such as a phase-stepping curve or beamlet/speckle) changes upon sample introduction. If there is a noisy pixel within that window,  or strong overlying contrast due to a phase or absorption edge, the image retrieval method can struggle, potentially leading to artefacts in the recovered images \cite{de2022high,lindberg2025kidney,Alloo2026DarkFieldFamily,croughan2024correcting}. Because methods derived from the XFPE can model and/or isolate phase edges \cite{alloo2025separating} and utilize more pixels in a given calculation, this minimizes such issues. 
\\\\
In this paper, we develop a novel XFPE-based dark-field retrieval method for GI. This is achieved by transforming GI data into a form comparable to single-grid/grating/speckle-based imaging data, and then applying an XFPE approach which we have developed for such data \cite{beltran2023fast,alloo2025separating}. Note that here, and in many cases, phase and dark-field retrieval algorithms can be interchangeably applied to speckle-based, single-grid, and single-grating datasets - a grid or grating is simply a speckle pattern with a single speckle size or modulation frequency \cite{morgan2012x,zdora2017x,beltran2023fast}. By interlacing GI data acquired at different analyzer grating positions, an image pair can be formed that resembles a reference 1D grating image and a sample-and-grating image, in which the grating pattern is directly resolved. This image pair can then be used as input to the relevant XFPE retrieval method to recover transmission and dark-field images, yielding two complementary images of the sample \cite{beltran2023fast,alloo2025separating}. To demonstrate our new approach, the remainder of this manuscript is arranged as follows. Section~\ref{Methods} will: \textit{A}, derive our new XFPE image retrieval method for GI, describing both the interlacing step and the solution of the inverse problem and \textit{B}, describe the experimental grating-interferometry imaging system used to acquire data from a test sample and a mouse chest to demonstrate our approach. In Sec.~\ref{Results}, we present and discuss the retrieved images for all samples using the XFPE method, and compare them qualitatively and quantitatively to those obtained using the conventional sinusoidal fitting approach \cite{Pfeiffer2008}. We finish with final remarks and potential avenues for future work.
\section{Methods}\label{Methods}
\begin{figure*}[htb] 
    \centering
    \includegraphics[width=\textwidth]{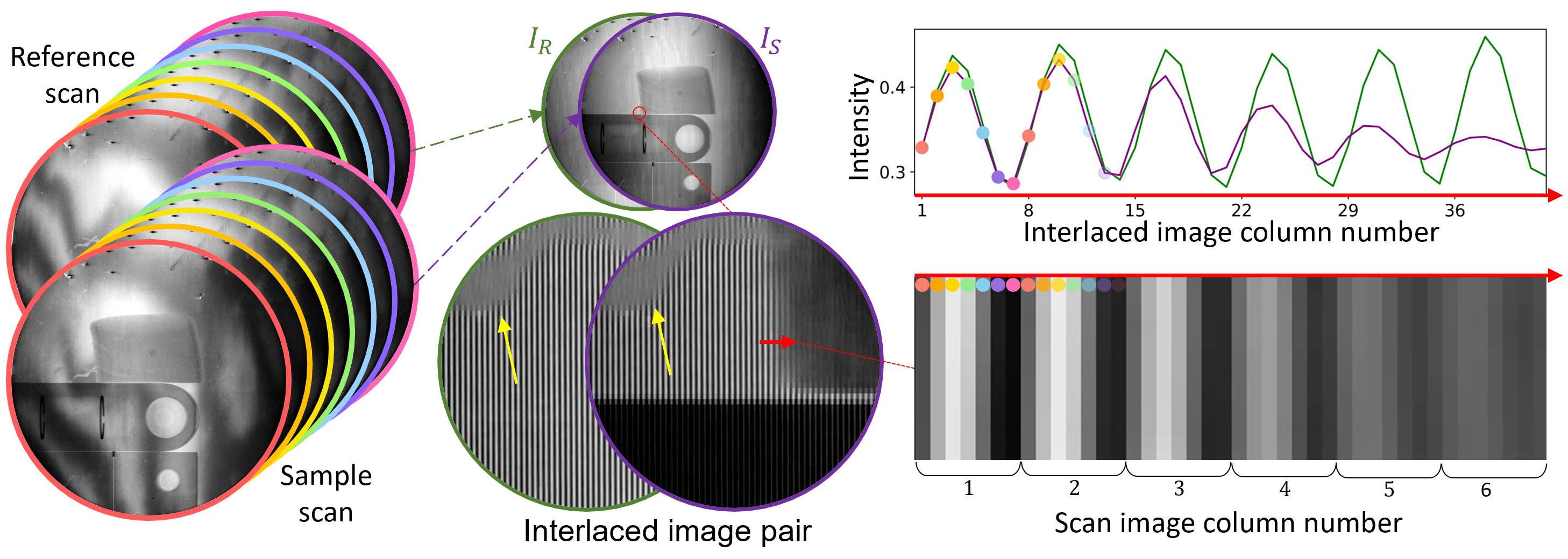}
    \caption{\textit{Overview of the interlacing procedure used to convert GI data into a form suitable for X-ray Fokker--Planck image retrieval. Columns from each scan image (7 different $G_2$ positions) are stacked sequentially: the first column from all 7 images side-by-side, then the second column, and so on, forming a single interlaced image. This image is resized to a square by duplicating each row 7 times until its height matches the width. The final interlaced images for the reference ($I_R$) and sample ($I_S$) scans are shown with green and purple borders, respectively. The magnified region and accompanying plot (taken across the red arrow in the magnified $I_S$ image) on the right illustrate that every 7 columns encode one phase-stepping series (sinusoidal reference pattern), which is modified by sample structure. The yellow arrows in the magnified interlaced image pair indicate a region of grating damage. }}
    \label{fig:Approach}
\end{figure*}

\subsection{Image Retrieval via the X-ray Fokker--Planck-Equation}\label{methodIR}
\noindent\textbf{Step 1}: \textit{Interlace grating interferometry phase-stepping scans}
\\In order to apply those XFPE image retrieval methods designed for speckle or single-grid data \cite{beltran2023fast,alloo2025separating} to GI data, that data must first be transformed into a suitable form where a reference pattern is visible. To do this, the experimental phase-stepping data are interlaced such that the intensity variations across $n$ analyzer grating steps for a given pixel are laid out across $n$ adjacent columns of a larger image, resulting in a bright and dark band reminiscent of a resolved grating line in a high-resolution single-grating image \cite{bennett2010grating,morgan2010quantitative}. As shown in Fig.~\ref{fig:Approach}, the reference scan is rearranged into an interlaced reference image, and the sample scan is rearranged into an interlaced sample image, each of which includes a fine, resolvable 1D periodic reference pattern. To construct the interlaced image, we take the first column of pixels from the image at the first analyzer grating position and place it into a larger empty array in the same position. Next, the first column of pixels from the image at the second analyzer position is appended adjacent to the first in the interlaced image, and so on, until the first column from all $n$ analyzer steps has been laid out and occupies the first $n$ columns of the interlaced image ($n=7$ in Fig. \ref{fig:Approach}). The same process is repeated for the second pixel column of the scan images, continuing iteratively until all columns from all analyzer positions have been included in the interlaced image, which will be $n$ times wider than the original image. We interlace the columns of the imaging data, not the rows, because sensitivity is known to be perpendicular to the grating lines (that is, across the image columns). Indeed, interlacing in the perpendicular direction and using it within an XFPE image retrieval algorithm yields images with artefacts and lower quality than those obtained from column-wise interlacing, and thus is not the recommended approach. Figure~\ref{fig:Approach} provides a summary illustration of the data interlacing procedure. Compared to the interlaced reference image (green), the sinusoidal pattern in the interlaced sample image (purple) will be modified in regions where the sample attenuates, refracts, or diffuses the X-rays. This is seen in the line profile plot in Fig.~\ref{fig:Approach}, which shows traces from the interlaced reference and sample images across the edge of a piece of wood in our test sample, with the strong dark-field signal from the wood significantly reducing the visibility or relative amplitude of the sinusoid.
\\\\
\textbf{Step 2}: \textit{Resize interlaced images}\\
The pair of interlaced images $I_R$ and $I_S$ will be $n$ times wider than the input experimental images. Since XFPE image retrieval relies on Fourier-space analysis, the interlaced images need to be resized vertically to preserve square pixels \cite{beltran2023fast, alloo2025separating}; we do this by duplicating each row of pixels $n$ times until its height matches its width; the resized interlaced images are shown in the center of Fig.~\ref{fig:Approach}. The resizing used row duplication, rather than interpolation, to enable a fair comparison between the Fokker--Planck approach and the conventional image retrieval approach, which does not use resizing.
\\\\
\textbf{Step 3}: \textit{Apply single-exposure XFPE image retrieval}\\
The generated pair of square, interlaced reference and sample images is of an appropriate form to be used within two pre-existing single-exposure XFPE algorithms for speckle/grid/grating-based X-ray imaging: Alloo \textit{et al.} \cite{alloo2025separating} and Beltran \textit{et al.} \cite{beltran2023fast}, which use the two distinct forms of the XFPE, the evolving and devolving, respectively. The evolving XFPE models the transformation of the reference pattern $I_R$ into $I_S$ when the sample is introduced, while the devolving XFPE models the reverse process occurring upon sample removal, that is, $I_S(\textbf{r}) \rightarrow I_R(\textbf{r})$. `Single-exposure' denotes that the retrieval algorithm uses a single image of the sample, opposed to methods that scan the reference pattern across the sample and use multiple sample exposures \cite{pavlov2020x}. 
\\\\
The finite-difference form of the evolving XFPE for speckle/grid/grating illumination is, 
\begin{equation}
I_S(\mathbf{r}) = I_R(\mathbf{r})\, t(\mathbf{r})
- \frac{\Delta}{k} \nabla_\perp \cdot
\left[ I_R(\mathbf{r})\, t(\mathbf{r}) \nabla_\perp \phi(\mathbf{r}) \right]
+ \Delta^2 \nabla_\perp^2
\left[ D(\mathbf{r})\, I_R(\mathbf{r})\, t(\mathbf{r}) \right],
\label{eqn:EvolveXFPE}
\end{equation}
and the physically equivalent devolving XFPE is
\begin{equation}
I_R(\mathbf{r}) = \frac{1}{t(\mathbf{r})} \left(
I_S(\mathbf{r})
+ \frac{\Delta}{k} \nabla_\perp \cdot
\left[ I_S(\mathbf{r}) \nabla_\perp \phi(\mathbf{r}) \right]
- \Delta^2 \nabla_\perp^2
\left[ D(\mathbf{r}) I_S(\mathbf{r}) \right]
\right).
\label{eqn:DevolveXFPE}
\end{equation}
In the above equations, $I_R(\textbf{r})$ denotes the reference image, containing a resolved speckle, grid, or grating pattern, $I_S(\textbf{r})$ the image collected when the sample is introduced, $\Delta$ is the sample-to-$G_2$ distance, $k$ is the X-ray wavenumber defined as $k = 2\pi/\lambda$ where $\lambda$ is the wavelength, $t(\textbf{r})$ is the transmitted intensity through the sample, $\phi(\textbf{r})$ is the phase shift induced by the sample, and $D(\textbf{r})$ is the XFPE effective diffusion coefficient, which characterizes the amount of local diffuse scattering, or dark-field effects, from the sample. The operators $\nabla_\perp$, $\nabla_\perp\cdot$, and $\nabla_\perp^2$ are the transverse gradient, divergence, and Laplacian operators, respectively, in the $\textbf{r} = (x,y)$ plane. The second term on the right-hand sides of Eqs.~\ref{eqn:EvolveXFPE} and~\ref{eqn:DevolveXFPE} describe how phase effects imposed by the sample modify the spatial distribution of the X-ray intensity through phase-gradient and Laplacian-phase effects, while the third term describes the local spread or diffusion of intensity due to sub-pixel X-ray scattering. The phase-retrieved transmission and dark-field images can be obtained from Eq.~\ref{eqn:EvolveXFPE} using the single-exposure approach of Alloo \textit{et al.} \cite{alloo2025separating}, and Eq.~\ref{eqn:DevolveXFPE} can be solved with a similar method developed prior by Beltran \textit{et al.} \cite{beltran2023fast}. Descriptions of both approaches are provided in Alloo \textit{et al.} \cite{alloo2025separating}, and readers are referred to that publication for details beyond what is provided below---refer to Secs.~3.1 and~3.2 in \cite{alloo2025separating} for the solutions for Eqs.~\ref{eqn:EvolveXFPE} and~\ref{eqn:DevolveXFPE}, respectively. 
\\\\
For both the evolving and devolving single-exposure XFPE image retrieval approaches, the phase-retrieved transmission image $t_\textrm{XFPE}(\textbf{r})$ is first reconstructed using the TIE-derived approach for speckle-based imaging \cite{pavlov2020single}:
\begin{equation}
t_\textrm{XFPE} \approx \mathcal{F}^{-1}\left[\frac{1}{1+\frac{\Delta\gamma}{2k}\left(k_x^2 +k_y^2\right)}\mathcal{F}\left[ \frac{I_S}{I_R}\right] \right].
\label{eqn:Trans}
\end{equation}
Above, $\mathcal{F}$ is the two-dimensional Fourier transform with respect to the transverse coordinates $x$ and $y$ perpendicular to the optical axis with $k_x$ and $k_y$ being the associated Fourier-space variables, and $\gamma=\delta/\beta\approx$ constant (the so-called homogeneous sample assumption), where $\delta$ and $\beta$ are the components of the sample's complex refractive index defined as $n=1-\delta+i\beta$. In Eq.~\ref{eqn:Trans}, we have dropped the explicit $(\textbf{r})$ dependence, the position in the image, for presentation clarity. Equation~\ref{eqn:Trans} approximates sample transmission as it neglects dark-field effects--this is reasonable when imaging samples with weak dark-field, where the experimental data is not significantly locally blurred. For samples that produce a strong dark-field, the transmission solution can be refined using XFPE to account for these effects; such an iterative scheme was recently presented by Liu et al. \cite{liu2025low}. However, this extension is not required in the present work, since any resulting artefacts will lie within a single period of the reference pattern, which is ultimately binned into a single pixel in the final step of the presented GI-XFPE method. Future work could explore whether an iterative approach could further improve the reconstructions beyond those presented here, which represent the first demonstration of XFPE-based image retrieval for GI.
\\\\
By substituting the retrieved $t_\textrm{XFPE}$ into the evolving and devolving XFPEs (Eqs.~\ref{eqn:EvolveXFPE} and~\ref{eqn:DevolveXFPE}, respectively) and rearranging, two realizations of the object's dark-field image can be retrieved from a single dataset. Note that this step uses the approximation that $\phi\approx-k\delta t_\textrm{XFPE}$. The evolving $D_{\textrm{Ev.}}(\textbf{r})$ and devolving $D_{\textrm{Dev.}}(\textbf{r})$ dark-field images can be retrieved using
\begin{equation}
\begin{aligned}
D_{\textrm{Ev.}} = \frac{1}{\Delta^2 I_R t_\textrm{XFPE}} \,
\nabla_\perp^{-2} \Big[
& I_S - I_R t_\textrm{XFPE} \\
& + \frac{\Delta \gamma}{2k} \Big(
  \partial_x \big[I_R t_\textrm{XFPE} \, \partial_x \ln(t_\textrm{XFPE})\big] 
  + \partial_y \big[I_R t_\textrm{XFPE} \, \partial_y \ln(t_\textrm{XFPE})\big]
\Big)
\Big]
\end{aligned}
\label{eqn:DFevolve}
\end{equation}
and
\begin{equation}
\begin{aligned}
D_{\textrm{Dev.}} = \frac{1}{\Delta^2 I_S} \,
\nabla_\perp^{-2} \Bigg[
& I_S - I_R t_\textrm{XFPE} \\
& + \frac{\Delta \gamma}{2k} \Big(
    \partial_x \big[I_S \, \partial_x \ln(t_\textrm{XFPE})\big] 
  + \partial_y \big[I_S \, \partial_y \ln(t_\textrm{XFPE})\big]
\Big)
\Bigg]
\end{aligned}
\label{eqn:DFdevolve}
\end{equation}
respectively. Above, $\partial_x$ and $\partial_y$ denote partial derivatives with respect to $x$ and $y$, respectively, which can be evaluated stably using $\partial_{x,\:y} = \mathcal{F}^{-1}ik_{x,\:y}\mathcal{F}$ via the Fourier derivative theorem \cite{paganin2006coherent}. While in this study we observed that the dark-field solutions given by Eqs.~\ref{eqn:DFevolve} and~\ref{eqn:DFdevolve} converge for GI image retrieval, Alloo \textit{et al.} \cite{alloo2025separating} showed that they are not equivalent in speckle imaging. Specifically, the dark-field solutions differed near sharp structures, such as sample edges (see Fig.~5 in Ref.~\cite{alloo2025separating}); these differences are not observed in the GI-retrieved dark-field images in this work. This may be because GI's coarser resolution averages out edge effects, such as propagation-based fringes, which remain detectable at the higher resolutions typical of speckle imaging. Differences between $D_{\textrm{Ev.}}$ and $D_{\textrm{Dev.}}$ for GI were observed primarily in the noise characteristics of these images; thus, we averaged these solutions (in Step 6) to improve the noise statistics and enhance the quality of the final retrieved dark-field image.\\\\
The inverse transverse Laplacian operator in XFPE algorithms (present here in Eqs.~\ref{eqn:DFevolve} and~\ref{eqn:DFdevolve}) is typically implemented using a Tikhonov-regularized result of the Fourier-derivative theorem:
\begin{equation}
\nabla_\perp^{-2} = -\mathcal{F}^{-1}\left\{\left[{k_x^2+k_y^2+\epsilon}\right]^{-1}\mathcal{F}\left[\:\right]\right\},
\label{eqn:InvLap}
\end{equation}
where the empty square brackets specify the operator's argument position \cite{paganin2006coherent,leatham2023x,ahlers2024x}. The parameter $\epsilon$ is a small, positive value chosen to stabilize the operator near the Fourier-space origin, where division by zero leads to ill-posed spatial frequencies. Recently, we developed an iterative method to optimize the value of $\epsilon$ for XFPE approaches that use multiple pairs of $I_R$, $I_S$ \cite{alloo2024stabilizing}. This method is not applicable when only a single set of imaging data is available, as in our XFPE implementation for GI here. Typically, in these single-dataset cases, the value of $\epsilon$ is chosen by manual inspection to ensure that the retrieved image is artefact-free and physically consistent. In this work, we provide an explicit expression for $\epsilon$ that accurately regularizes $\nabla_\perp^{-2}$ for a given imaging dataset based on the measured signal-to-noise ratio (SNR) and imaging system's pixel size $p$, thereby removing the need for manual tuning typically required in single-exposure XFPE methods:
\begin{equation}
\epsilon = \frac{1}{\textrm{SNR}\:p^2}=\frac{\sigma_\textrm{noise}}{\mu_\textrm{signal}\:p^2}.
\label{eqn:OptimalReg}
\end{equation}
Above, the SNR is defined as the mean intensity $\mu_\textrm{signal}$ of a region enclosing the sample (that is, from an image in the sample scan) divided by the standard deviation of the intensity $\sigma_\textrm{noise}$ of a region containing only air. Intuitively, the equation is sensible for the following two reasons; (1) data that has a lot of noise requires stronger regularization, consistent with Eq.~\ref{eqn:OptimalReg} being inversely proportional to SNR, and (2) dimensionally, the regularization parameter has units of inverse length squared, which is given by the inverse proportionality to pixel size in Eq.~\ref{eqn:OptimalReg}. A complete derivation and explanation of Eq.~\ref{eqn:OptimalReg} are provided in Sec.~1 of the Supplementary Material.
\\\\
\textbf{Step 4}: \textit{Returning the retrieved images to their original size}\\
Applying the single-exposure XFPE methods described in Step 3 to the interlaced images (see Fig. \ref{fig:Approach}) retrieves transmission and dark-field images that are $n$ times larger than the imaging data, where $n$ is the number of analyzer grating positions in the scan. Each $n\times n$ block in the retrieved image represents the information gathered by one pixel of the GI scan ($n$ grating steps from the one pixel, horizontally, replicated over $n$ steps vertically for Fourier space calculations), so no additional resolution is represented in these additional pixels. In addition, we seek to compare the result to conventional methods over the same pixel array. It therefore makes the most sense to average each $n\times n$ block into one pixel to maximize SNR, yielding a (i) phase-retrieved transmission image, (ii) evolving-retrieved dark-field image, and (iii) devolving-retrieved dark-field image within the original dimensions of the GI data.
\\\\
\textbf{Step 5}: \textit{Repeat steps 1--4 with `reverse' interlacing}
\\
Because the grating steps correspond to the same pixel, the interlacing order in Step 1 is not fixed and can be reversed. A second interlacing can therefore be performed, placing the grating steps from right to left instead of left to right, allowing a second phase-retrieved transmission image and dark-field image to be reconstructed (similar to how the devolving and evolving XFPE models do), enabling further averaging (Step 6) to improve the SNR of the final retrieved images. 
\\\\
\textbf{Step 6}: \textit{Average to retrieve the final optimal images}\\
Completing Steps 1--5 retrieves six images:
\begin{itemize}[noitemsep, topsep=0pt]
    \item Two phase-retrieved transmission images, one using each interlacing method,
    \item Four dark-field images, two using the evolving XFPE approach and two using the devolving XFPE approach, where the two come from each of the interlacing methods.
\end{itemize}
The two $t_\textrm{XFPE}$ solutions are averaged to obtain the final phase-retrieved transmission image, and the four $D_\textrm{XFPE}$ solutions are averaged to obtain the final dark-field image. Our XFPE approach can be considered an ensemble method, where multiple independent solutions from the same dataset are computed and then combined via averaging to produce a superior aggregate result in terms of image quality. Analysis revealed that combining these solutions improved the retrieved images' SNR and sharpness relative to a single reconstruction alone, improving our XFPE retrieval method's overall performance. 
\subsection{Imaging Experiments}\label{experiments} We apply our XFPE algorithm on GI data from two samples: a test sample comprising wood, tissue paper, and a spirit level (shown in the schematic of the GI setup in Fig.~\ref{fig:Setup}), and then a mouse chest. 
\subsubsection{Test sample}
The X-ray source used to image the test sample was a commercially available X-ray tube, the XWT-160-SE from X-RAY WorX GmbH (Garbsen, Germany), an open microfocus tube operating in reflection mode with a Tungsten target. It was operated at a 60 kVp, with a $\SI{2}{\milli\meter}$ Al window providing a polychromatic spectrum with a mean energy of around 45 keV. The GI setup (shown in Fig.~\ref{fig:Setup}) was designed for 45 keV X-rays and included a gold source grating $G_0$ fabricated on a silicon substrate, positioned 35 mm upstream of the source exit window. The gold structures were approximately 160--$\SI{170}{\micro\meter}$ in height, arranged in a periodic pattern with a $\SI{10}{\micro\meter}$ pitch, and 0.5 duty cycle. The phase grating $G_1$ was placed $\SI{925}{\milli\meter}$ downstream of $G_0$, it had a period of $\SI{5}{\micro\meter}$, a duty cycle of 0.5, and consisted of nickel bars $\SI{8}{\micro\meter}$ in height on a silicon substrate, providing a $\pi$/2 phase-shift at the 45 keV design energy. The sample was positioned $\SI{241}{\milli\meter}$ downstream from $G_1$, and the analyzer grating $G_2$ was positioned $\SI{684}{\milli\meter}$ downstream of the sample (in the first fractional Talbot distance $\SI{925}{\milli\meter}$ downstream of $G_1$). $G_2$ had a period of $\SI{10}{\micro\meter}$ with a duty cycle of 0.5 and consisted of 160--$\SI{170}{\micro\meter}$ high bars of gold on a silicon substrate. The detector, positioned directly downstream of $G_2$, was a flat panel Varex Imaging XRD 4343RF (Salt Lake City, UT, USA) with a CsI scintillator, which had a $\SI{150}{\micro\meter}$ pixel pitch and a 2800 $\times$ 2880 pixel matrix. Taking into account the GI setup's 1.7$\times$ sample magnification, the effective pixel size and propagation distance (taken as the sample-to-$G_2$ distance) were $\SI{88.23}{\micro\meter}$ and $\SI{40.23}{\centi\meter}$, respectively. To acquire the GI imaging dataset, images were taken at seven $G_1$ positions, evenly spaced over one grating period; a reference scan without the sample in the beam was first performed, and the sample was then placed into the interferometer to acquire the sample scan. Each image in the phase-stepping series was acquired using a 1 s exposure time. Due to hardware limitations, the stepping positions during data acquisition were not perfectly evenly spaced. We used the algorithm by Hashimoto \textit{et al.} \cite{hashimoto2020improved} to extract exact stepping positions, retrieved the image parameters, and re-created perfectly evenly-spaced intensity data.

\subsubsection{Mouse chest}\label{mousechest_imaging}
Imaging of a mouse chest was performed at the Munich Compact Light Source (MuCLS) at the Technical University of Munich (TUM), a source based on inverse Compton scattering for highly brilliant quasi-monochromatic X-ray generation \cite{gunther2020versatile}. This dataset was first published in Gradl \textit{et al.}~\cite{gradl2018dynamic} and further details on the source and experimental setup beyond what is given here can be found there. A partially coherent 25 keV monochromatic energy X-ray beam was used for imaging, meaning that a source grating was not required in this setup. A GI setup was established using a phase grating $G_1$ placed $\SI{15}{\meter}$ from the X-ray source. The phase grating $G_1$ had nickel grating lines produced with a height of $\SI{4.39}{\micro\meter}$, inducing a phase shift of $\pi$/2 at the design energy of 25 keV. The period of $G_1$ was $\SI{4.92}{\micro\meter}$ and it had a duty cycle of 0.5. The sample, that is, the mouse, was located $\SI{130}{\milli\meter}$ downstream of $G_1$. The analyzer grating $G_2$ was placed at the first fractional Talbot distance of $\SI{248}{\milli\meter}$ downstream of $G_1$. The grating bars in $G_2$ were made from gold and were $\SI{70}{\micro\meter}$ in height, with a period of $\SI{5}{\micro\meter}$, and a duty cycle of 0.5. Placed directly behind the analyzer grating $G_2$ was a Pilatus 200K detector (Dectris Ltd., Baden, Switzerland) with a $\SI{1000}{\micro\meter}$ thick silicon sensor. The detector had 487$\times$407 pixels, each with a $\SI{172}{\micro\meter}$ pitch. The magnification of the GI setup was 1.0078, giving an effective pixel size and propagation distance of $\SI{170.65}{\micro\meter}$ and $\SI{11.91}{\centi\meter}$. The mouse was imaged dynamically, meaning that it was ventilated over an 84 s imaging period, and images were acquired at different points over one breath for 7 breaths. A detailed description of the procedure used for dynamic imaging can be found in Gradl \textit{et al.}~\cite{gradl2018dynamic} (see Fig.~1 therein). All animal procedures were conducted in accordance with protocols approved by the Regierung von Oberbayern (District Government of Upper Bavaria). Mice were housed in individually ventilated cages (IVC racks; BioZone, Margate) with filtered air under a 12 h light/12 h dark cycle (lights on from 06:00 to 18:00). Standard chow and water were provided ad libitum. Reference and sample scans of the mouse chest were acquired at exposure times of 40 ms, 90 ms, and 140 ms to explore the influence of noise on reconstruction quality. As a total imaging time of 84 s was used for all exposure times, the number of images acquired across one breath varied with exposure time. For 40 ms, 90 ms, and 140 ms, 30, 15, and 10 images per breath were recorded, respectively, for a total of 7 breaths. For each dataset, we selected the image acquired at the top of inhalation, that is, when the lungs were the largest in volume, to present in this manuscript. The raw images at this breath point were averaged over the 7 breaths recorded for each of the 7 positions of the analyzer grating, compiling the sample scan. Although only static image results are presented in the main manuscript, the corresponding time-resolved reconstructions for the dynamic study are provided as videos in an open Zenodo repository \cite{alloo2026zenodo}. Because the exposure time was not sufficiently short to show visible noise in the reconstructed images, we therefore simulated an even shorter exposure and examined the performance of the XFPE image-retrieval method on that dataset. This was simulated by first scaling image intensities by a factor of 1/2 to simulate reduced photon statistics, corresponding to an effective exposure time of 20 ms. A pixel-wise Poisson random process was then applied to the sample scan images (the only images subject to dose/exposure constraints), with the scaled intensity used as the rate parameter, and the resulting counts converted back to floating-point values.
\section{Results}\label{Results}
To demonstrate and validate our GI-XFPE approach, we compare the XFPE-retrieved images with those obtained using pixel-wise sinusoidal fitting (hereafter referred to as the `conventional' method) \cite{Pfeiffer2008}, for both samples. Readers are referred to Ref.~\cite{Pfeiffer2008} for details of the conventional method.
\begin{figure}[htb!] 
    \centering
    \includegraphics[width=0.6\textwidth]{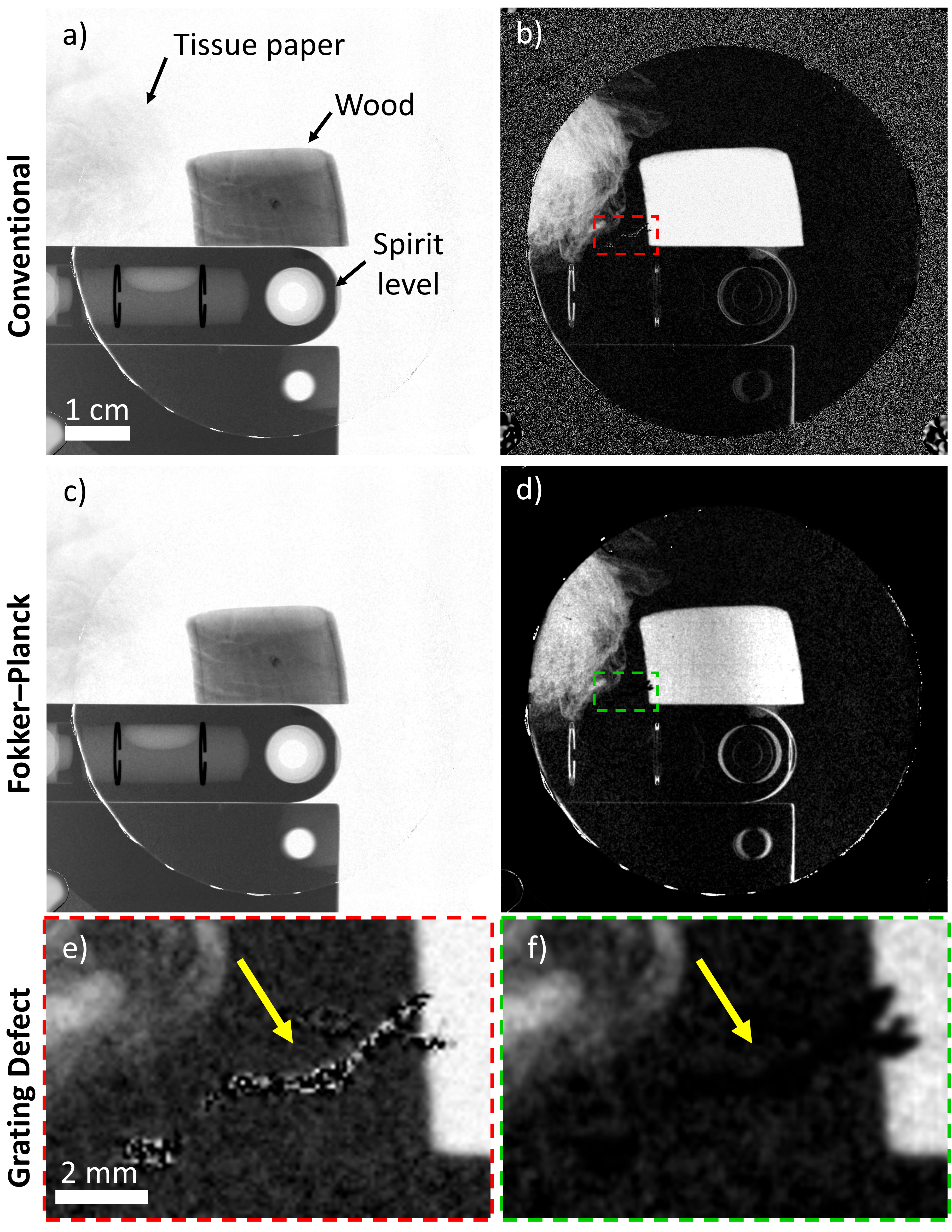}
    \caption{\textit{Retrieved (a) transmission and (b) dark-field images of the test sample using the conventional pixel-wise sinusoidal fitting method. Retrieved (c) transmission and (d) dark-field images using our new XFPE-based approach. (e) and (f) are a magnified region from both dark-field images, highlighting how a grating defect--indicated by the yellow arrow in c--is handled by the two methods. The conventional approach yields a rapidly varying dark-field signal at the defect location, whereas the XFPE method retrieves no signal in this region. Images are displayed on a linear grayscale of [min (black), max (white)]: a, c) [0.6, 1.0], b, e) [0.0, 1.0] (1-visibility change), d, f) [0.0, 1.5$\times10^{-9}$] ($D$)}.
    }
    \label{fig:TestSam}
\end{figure}
\subsection{Test Sample}
Figure~\ref{fig:TestSam} shows the retrieved transmission and dark-field images using the conventional and XFPE methods. The transmission images are displayed on a common grayscale range, whereas the dark-field images use different scaling, as the conventional method retrieves a visibility reduction while the XFPE method retrieves the Fokker--Planck effective diffusion coefficient. The display ranges were chosen to ensure comparable dark-field contrast across materials in both cases. We note that, while the conventional method is also sensitive to the differential phase-contrast signal, we only show transmission results in this paper, as this XFPE method couples the retrieved phase and transmission signals, meaning these images are not retrieved independently. Note, that for the GI setups considered here, which utilize relatively large pixels and short sample-to-detector propagation distances compared to speckle-based imaging \cite{alloo2023m,beltran2023fast}, propagation-based phase-contrast fringes are not resolved. This means the phase-retrieval filter in real space is less than one pixel wide, so Eq.~\ref{eqn:Trans} reduces to $I_S$/$I_R$. 
\\\\
The XFPE-retrieved transmission image in Fig.~\ref{fig:TestSam}c agrees qualitatively and quantitatively with the conventional result in Fig.~\ref{fig:TestSam}a, whereas the dark-field images exhibit differences. From visual inspection, structures expected to generate dark-field, such as the tissue paper and the block of wood, are consistently retrieved via both XFPE and conventional methods. Subtle differences in the dark-field appear at sample edges, in the background noise characteristics (air-filled regions), and at locations corresponding to defects in the experimental data. The edge-generated dark-field \cite{yashiro2010origin,yashiro2015effects,alloo2025separating} is stronger in the XFPE image than in the conventional method, seen, for example, at the edges of the spirit level in Figs.~\ref{fig:TestSam}b and~\ref{fig:TestSam}d. Moreover, in the XFPE-retrieved dark-field (Fig.~\ref{fig:TestSam}d), the edge-generated signal is comparable in strength to the dark-field arising from microstructure-dense regions, such as the tissue and wood, whereas the edge dark-field signal is significantly weaker in the conventionally retrieved dark-field (Fig.~\ref{fig:TestSam}b). Another clear difference is how the two methods handle defects in the raw experimental data, as highlighted in the magnified images in Figs.~\ref{fig:TestSam}e and~\ref{fig:TestSam}f. The G1 grating used in the GI system was damaged in a small region, with the position of the scratch indicated by the yellow arrow in Figs.~\ref{fig:TestSam}e and ~\ref{fig:TestSam}f, leading to a local loss of usable signal. In the experimental data, available online \cite{alloo2026zenodo} and also shown in Fig.~\ref{fig:Approach} (yellow arrows), this region does not exhibit the intensity oscillations required for image retrieval. Via conventional dark-field retrieval, the lack of usable intensity modulation leads to random dark-field values, as shown in Fig.~\ref{fig:TestSam}e. A negligable dark-field signal is retrieved at the location of this grating defect via the XFPE approach, as shown in Fig.~\ref{fig:TestSam}f. We further notice that when the grating defect overlaps a dark-field-generating feature, such as the piece of wood, the XFPE approach returns this region as ``dark-field free", as the information in this area is effectively lost. 
\begin{figure*}[htb!] 
    \centering
    \includegraphics[width=\textwidth]{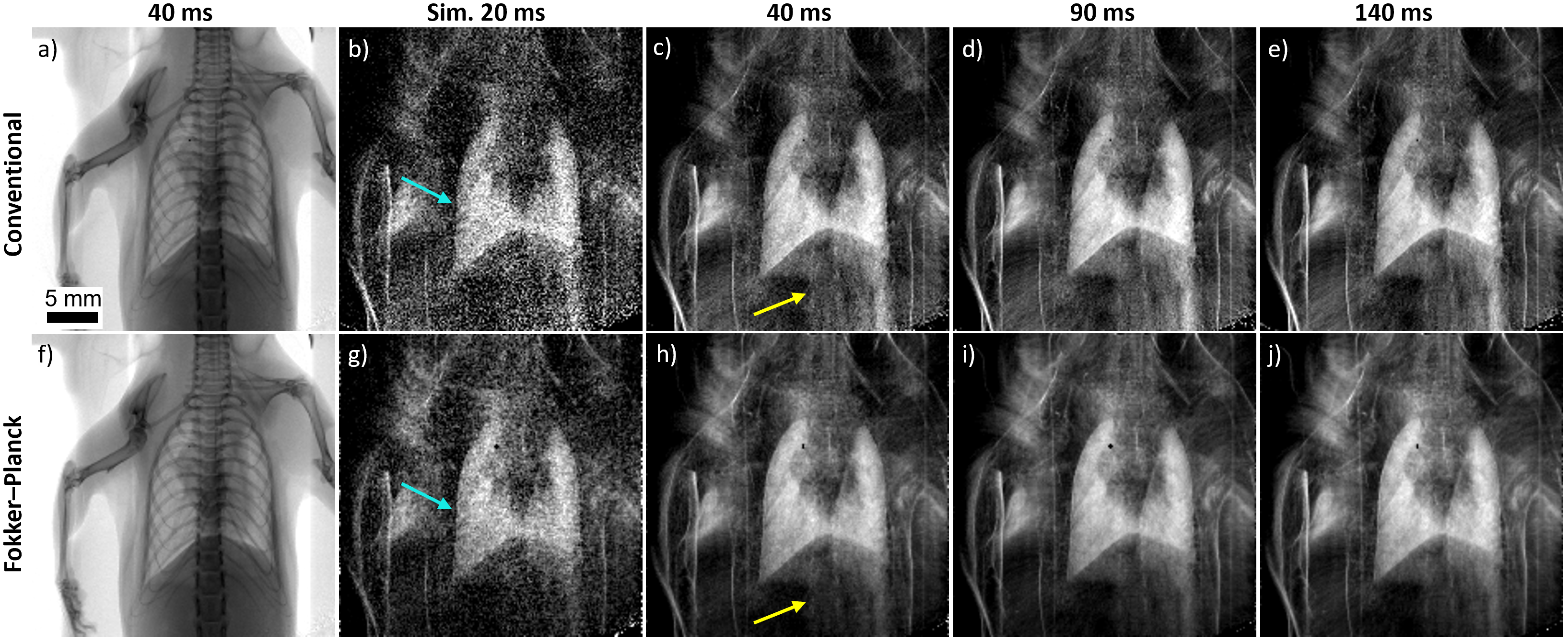}
    \caption{\textit{Retrieved transmission (1st column) and dark-field images (2nd--5th columns) of a mouse chest using the conventional pixel-wise sinusoidal fitting method (top row) and our new XFPE approach (bottom row). Results are shown for different exposure times, indicated along the top of the figure. The turquoise arrows in panels b) and g) indicate the lung edge, where the conventional retrieval method amplifies noise during processing, causing this boundary to appear less sharp than in the XFPE-retrieved dark-field image. The yellow arrows in panels c) and h) indicate a region where the conventional image-retrieval approach reconstructs significantly more noise (across all datasets in this figure) than the XFPE approach. Images are displayed on a linear grayscale of [min (black), max (white)]: a), f) [0.25, 1.0], b)--e) [0.1, 0.8], and g)--j) [1.0, 8.5]$\times10^{-9}$.}}
    \label{fig:Mouse}
\end{figure*}
\begin{figure}[tb] 
    \centering
\includegraphics[
  width=0.7\textwidth,
]{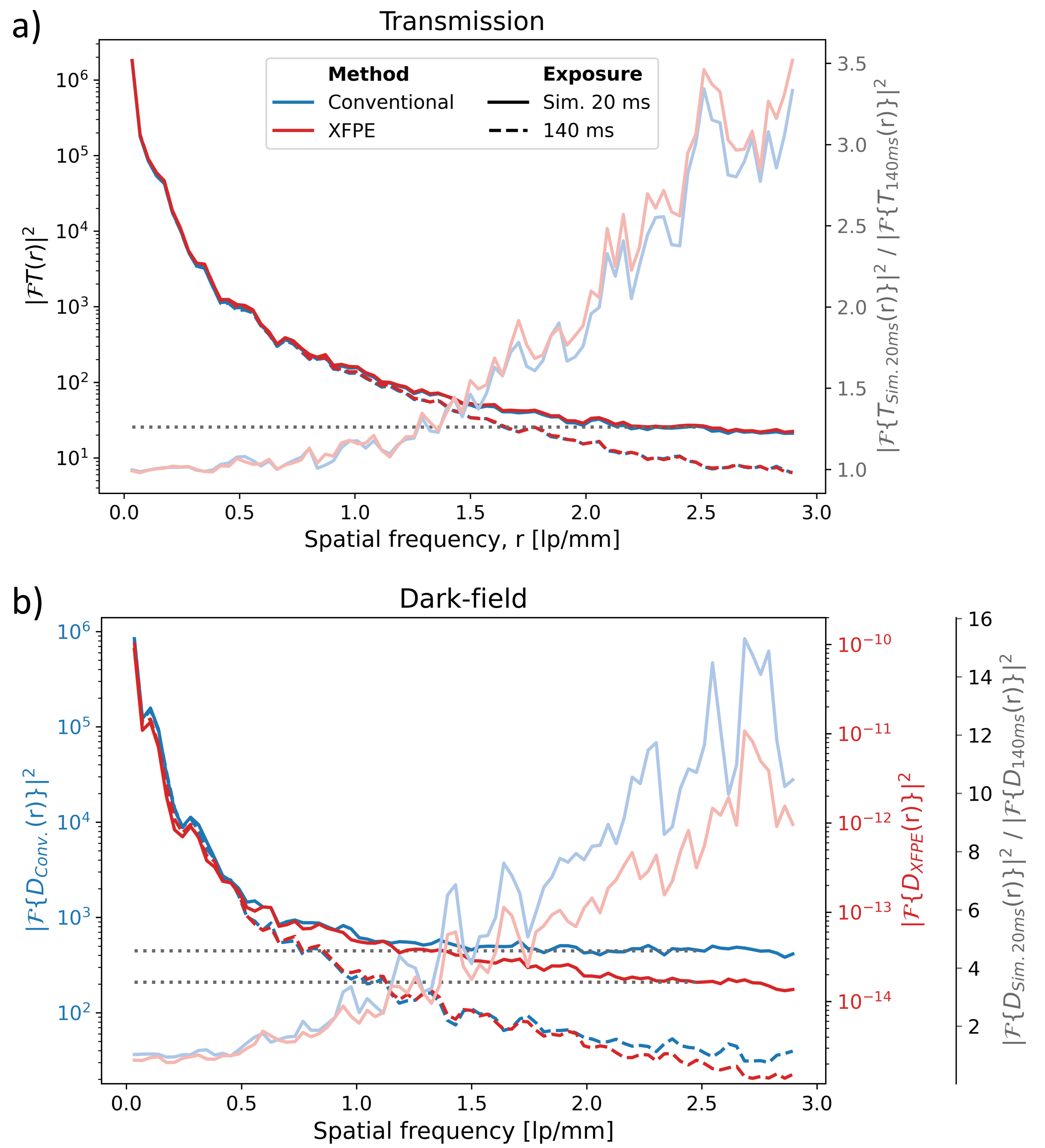}
    \caption{\textit{Azimuthally averaged Fourier power spectra of the retrieved a) transmission and b) dark-field images from simulated 20 ms (solid full-color curves) and experimental 140 ms (dashed full-color curves) grating interferometry (GI) data of the mouse chest. The transmission spectra are plotted on the same axis, whereas the dark-field spectra are displayed on separate axes because the approaches retrieve different dark-field metrics. The dotted grey lines on the simulated 20 ms dark-field spectra denote the approximate location of the noise floor. Semi-transparent blue and red curves represent the ratio of the 20 ms to the 140 ms spectra for the conventional and XFPE retrieval methods, respectively. The gray vertical axes in both plots denote the scale corresponding to these ratio curves.}}
    \label{fig:FPS}
\end{figure}
\subsection{Mouse Chest}
Dark-field X-ray imaging using GI has shown particular promise for clinical diagnostic chest imaging\cite{Willer2021,gassert2025dark,gassert2025comparison}. To assess the performance of XFPE image retrieval for this application, we compare image retrieval using the proposed XFPE method with that using a conventional method for the mouse chest data described in Sec.~\ref {mousechest_imaging}. Some of the retrieved images are shown in Fig.~\ref{fig:Mouse}. Here, we present the retrieved transmission images only for a single breath point and exposure time (40 ms), as the two image retrieval approaches yield similar transmission images across all exposure times. This similarity is quantitatively demonstrated in Sec.~2 of the Supplementary Material, where image quality measurements of the retrieved images obtained using the conventional and XFPE retrieval methods are presented (see Fig.~2 in the Supplementary Material). Further verification can be achieved by examining the retrieved transmission images available in Ref.~\cite{alloo2026zenodo}. As observed with the test sample, differences are more apparent in the retrieved dark field. Consistently, across all exposure times (that is, columns of Fig.~\ref{fig:Mouse}) but most apparent for the simulated 20 ms dataset, high-frequency noise is more visually apparent in the conventionally retrieved dark-field image, whereas the XFPE dark-field images are visually smoother. In the stomach/lower spine region of the mouse, denoted with the yellow arrows in Figs.~\ref{fig:Mouse}c and~\ref{fig:Mouse}h, the conventionally retrieved dark-field images exhibit dark-field contrast that is not present in the XFPE-retrieved dark-field images, and which would not be expected from this part of the anatomy. In this region, the transmitted flux in the sample scan is reduced to $\approx36\%$ due to absorption by the spine and stomach, meaning the sample stepping curve has fewer counts and more noise than other regions of the image. In addition, the mean value of the reference scan varies pixel-to-pixel in this area, with a standard deviation nearly three times that measured in the center of the image. Under strong attenuation and the resulting high-noise conditions, conventional processing exhibits a bias in the sinusoidal fit, which leads to an apparent dark-field signal in Figs.~\ref{fig:Mouse}b--\ref{fig:Mouse}e \cite{chabior2011signal,ji2017studies}. The different signal retrieval strategy used in the XFPE framework suppresses this artifact (Figs.~\ref{fig:Mouse}g--\ref{fig:Mouse}j).
\\\\
The mutual consistency between XFPE-retrieved and conventionally retrieved images, with the largest differences observed in the spatial frequency composition of the dark-field images, is further supported by the comparison of the azimuthally averaged Fourier power spectra, shown in Fig.~\ref{fig:FPS}. Note that, for clarity, only the spectra of the simulated 20 ms dataset and the experimental 140 ms dataset are shown--the spectra of the images from the 90 ms and 140 ms datasets are similar to those of the 140 ms. The similar Fourier power spectra of the retrieved transmission images for both exposure times, and the ratio between the two exposure times, shown in Fig.~\ref{fig:FPS}a, demonstrate the mutual consistency of the two retrieval approaches. As expected, the transmission images retrieved from the shorter-exposure data exhibit increased high-spatial-frequency content (in other words, a higher noise floor) compared with those from the longer-exposure data. For the dark-field images, the low- to mid-spatial-frequency structures are retrieved similarly using the conventional and XFPE methods, with the spectra largely overlapping in this region--shown in Fig.~\ref{fig:FPS}b (e.g. up to 2 lp/mm for the 140 ms exposures). At higher spatial frequencies, however, the spectra diverge, with the higher spatial frequencies consistently higher in the conventionally retrieved dark-field images than in the XFPE-retrieved images. This difference is more pronounced for the shorter-exposure dataset, where the blue curve sits above the red for the solid curves (20 ms), while this difference is slightly reduced for the dashed curves (140 ms) in Fig.~\ref{fig:FPS}b. These observations demonstrate that high-spatial-frequency information in the input data is handled differently by the conventional and XFPE image retrieval methods, particularly at short exposures where the data contain increased high-frequency noise. In the resulting dark-field images, the conventional method amplifies these high frequencies more strongly than the XFPE method, leading to a noisier image appearance. This conclusion is supported by two observations. (1) The images in Fig.~\ref{fig:Mouse} show a smoother dark-field signal from the mouse lungs for the XFPE approach compared to the conventional approach, which exhibits pronounced pixel-to-pixel fluctuations. However, this smoothing is not at the expense of spatial resolution - the turquoise arrows in Figs.~\ref{fig:Mouse}b and~\ref{fig:Mouse}g indicate the edge of the mouse lung, which appears sharper in the XFPE-retrieved image due to the reduced contribution of high-spatial-frequency noise. (2) The semi-transparent curves in Fig.~\ref{fig:FPS}b show the ratio of the Fourier power spectra of the short-exposure to long-exposure dark-field images for each method. The ratio curve for the conventional approach (blue) lies above that of the XFPE approach (red), particularly at high spatial frequencies, demonstrating that the conventional method amplifies these frequencies more than the Fokker-Planck approach when noise is introduced to the raw data. These observations are also consistent with the spatial resolution measurements obtained using the method of Modregger \textit{et al.} \cite{modregger2007spatial} (given in Sec.~2 of the Supplementary Material), which indicate the spatial resolution improves by a factor of 1.77 when using XFPE instead of conventional retrieval for the simulated 20 ms dataset, and 1.16 for the 140 ms exposure dataset. The high-spatial-frequency noise in the conventionally retrieved dark-field images leads to a degradation in the apparent spatial resolution computed using the Modregger \textit{et al.} metric \cite{gureyev2020noise}.
\\\\
Our results reveal that the XFPE approach shows an advantage over the conventional method for dark-field retrieval under short sample exposures and/or high Poisson noise. When repeating the 20 ms--Poisson noise simulation ten times, we found that the standard deviation in the retrieved dark-field value at a given pixel was, on average 17.3\% using the conventional retrieval approach and 10.7\% using XFPE, demonstrating the XFPE approach's robustness to noise (see Sec.~3 in the Supplementary Material). This result, the provided supplementary videos \cite{alloo2026zenodo}, together with the quantitative image-quality analysis presented here, indicate that the presented XFPE image-retrieval method for grating-interferometry-based chest imaging can improve the quality of the retrieved images, particularly in the dark-field channel.
\section{Future Work}\label{FW}
In this manuscript, we demonstrated the qualitative and quantitative effectiveness of XFPE-based image retrieval on GI data. This fundamentally new approach to retrieving transmission and dark-field images from the well-developed imaging setup opens several avenues for future research. \\\\
A key step in future work will be linking the physical interpretation of visibility reduction in GI \cite{bech2010quantitative,lynch2011interpretation,strobl2014general,prade2016short} with the XFPE-derived diffusion coefficient \cite{paganin2023paraxial, leatham2024x}, combining scattering theory and experimental measurements of samples with well-defined microstructure sizes across a range of experimental configurations. Investigating the relationship between these metrics may provide a pathway toward a physical interpretation of the newly introduced Fokker--Planck effective diffusion coefficient, and benefit those imaging techniques that utilize the XFPE for image retrieval, including propagation-based \cite{leatham2023x,ahlers2024x} and speckle-based \cite{pavlov2020x} X-ray imaging.
\\\\
With major efforts focused on applying GI as a diagnostic imaging tool \cite{Momose2014,Willer2021,Viermetz2022,Urban2023}, the system's sensitivity and robustness to both noise and vibrations are important \cite{pereira2025quantifying,meyer2025moire,haeusele2023advanced}. While the XFPE approach has been shown to be resilient to noise, future work could assess the performance of the proposed XFPE-based method relative to conventional image-retrieval approaches under other non-ideal conditions. Because the presented XFPE approach does not require a perfectly sinusoidal curve--that is, random speckles can be tracked--it may be helpful in cases where the fringe signal contains higher-order harmonic content or deviates from an ideal sinusoid. It would also be interesting to investigate algorithm performance and potential adjustments in the case of mismatches between reference and sample scans.
\\\\
The interlacing approach shown here, which enables XFPE-based image retrieval in GI, is expected to be applicable to other phase- and dark-field imaging setups that employ stepping, such as analyzer-based imaging \cite{forster1980double} and edge-illumination \cite{olivo2001innovative}. These techniques involve stepping an optical element (a crystal in analyzer-based imaging and a grating in edge-illumination) and acquiring reference and sample images at each step, in a manner analogous to GI. It is expected that the presented XFPE method could offer similar advantages for these techniques to those demonstrated in this work and in our previous studies, particularly in terms of noise resilience and improved spatial resolution. Establishing a cohesive image-retrieval framework across different imaging techniques is particularly advantageous, as it enables seamless quantitative comparison between methods, yielding identical retrieved values across approaches. The XFPE diffusion coefficient presents an advantage in this aspect because it is setup-independent \cite{morgan2019applying}. 
\\\\
An important difference between the presented XFPE image retrieval algorithm and the conventional sinusoidal-fitting algorithm lies in their ability to independently retrieve transmission, phase, and dark-field information at each pixel. The XFPE retrieval algorithm yields a closed-form analytical solution for the transmission and the dark-field, with pixels sharing information. Although the XFPE-retrieved transmission retrieval utilizes the homogeneous-sample assumption ($\gamma=\delta/\beta\approx$ constant) in theory, this is not a limitation in practice because GI is typically performed at a spatial resolution where no propagation-based fringes are visible and the phase retrieval filter (Eqn. \ref{eqn:Trans}) is hence very weak. In any case, numerous studies have demonstrated successful image retrieval while retaining this homogeneous-sample assumption across a wide variety of samples (see, for example, the third paragraph of Ref.~\cite{paganin2020boosting}). Future work could explore how transmission and differential phase signals in GI can be uniquely separated using an XFPE-based approach. 

\section{Conclusion}
In this work, we developed and validated a fundamentally new dark-field image retrieval algorithm for grating-interferometry data based on the X-ray Fokker--Planck equation. The XFPE-based algorithm interlaces a complete grating-interferometry dataset along the phase-stepping curves to generate two images that map all information acquired during the scan: an interlaced reference image and an interlaced sample image. Whole-image XFPE-based processing, using an algorithm similar to that developed for speckle-based X-ray imaging, is then performed to retrieve complementary transmission and dark-field images that aid visualization of dense and sub-resolution structures, respectively. We compared the developed grating-interferometry Fokker--Planck algorithm to the conventional pixel-wise sinusoidal fitting approach using a test sample comprised of known materials and a mouse-chest dataset (static and dynamic results presented). XFPE dark-field retrieval produces images that are consistent with conventional retrieval, with a comparable SNR. Differences between the retrieved images are observed where the grating pattern is altered, either by a scratch in the pattern (Figs.~\ref{fig:TestSam}e and~\ref{fig:Mouse}f), reduced flux (Fig.~\ref{fig:Mouse}, yellow arrow), reduced grating visibility (seen in the most extreme case by looking at the square area outside the round grating pattern, Fig.~\ref{fig:TestSam}), or with variations in Poisson noise (see Sec.~2 in the Supplementary Material and the supplementary videos accessible via Zenodo \cite{alloo2026zenodo}). The proposed Fokker--Planck method offers an advantage over conventional dark-field retrieval for fast sample imaging under short exposure times and high noise, yielding improved image quality. The conventional approach operates on a single-pixel basis and is therefore more vulnerable to these issues, which can result in high-frequency noise. In contrast, the X-ray Fokker--Planck equation approach incorporates information from neighboring pixels, making it less sensitive to such effects, yielding dark-field values closer to zero and more consistent results across exposures. To encourage adoption and integration, we publish our test data, alongside Python code implementing the developed image retrieval method, in an open Zenodo repository, with the aim of supporting grating-interferometry image-processing pipelines under development worldwide.

\section*{Acknowledgment}
The authors acknowledge useful discussions with David M. Paganin and thank Martin Dierolf for his assistance in acquiring the grating-interferometry data of the mouse chest at the Munich Compact Light Source (MuCLS). Kaye S. Morgan acknowledges support from the Australian Research Council (FT18010037 and DP230101327), Samantha J.~Alloo acknowledges support from an AINSE Ltd.~Early Career Researcher Grant (ECRG), and Benedikt G\"unther acknowledges support by the Deutsche Forschungsgemeinschaft, DFG, German Research Foundation (513827659).

\bibliographystyle{unsrt}
\bibliography{references_notitle}

\end{document}